\documentstyle[preprint,aps,epsf]{revtex}                  
\input epsf.tex
\def\DESepsf(#1 width #2){\epsfxsize=#2 \epsfbox{#1}}
\global\let\epsfloaded=Y 
\begin{document}
\pagestyle{empty}                                      
\preprint{
\font\fortssbx=cmssbx10 scaled \magstep2
\hbox to \hsize{
\hfill $
\vtop{
 \hbox{ }}$
}
}
\draft
\vfill
\title{Triple Neutral Gauge Boson Couplings\\ in Noncommutative Standard Model}
\vfill
\author{$^1$N.G. Deshpande and $^2$Xiao-Gang He}
\address{\rm $^1$ Institute of Theoretical Sciences, 
University of Oregon, Eugene, 
OR97403, USA\\
and\\
\rm $^2$Department of Physics, National Taiwan University,
Taipei, Taiwan 10764, R.O.C.}

%
%
\vfill
\maketitle
\begin{abstract}
It has been shown recently that the triple neutral gauge boson
couplings are not uniquely determined in noncommutative extension
of the Standard Model (NCSM). Depending on specific schemes used, the couplings
are different and may even be zero. 
To distinguish different realizations of the
NCSM, additional information either from
theoretical or experimental considerations is needed. In this paper we 
show that these couplings can be uniquely determined from 
considerations of unification of electroweak and strong interactions.
Using $SU(5)$ as the underlying theory and integrating out the heavy 
degrees of freedom,
we obtain unique non-zero new triple $\gamma\gamma\gamma$, $\gamma \gamma Z$, 
$\gamma ZZ$, $ZZZ$,
$\gamma GG$, $ZGG$ and $GGG$ couplings at the leading order in the NCSM. 
We also briefly discuss experimental implications.
 
\end{abstract}
%
%
\pacs{PACS numbers:
 }
%
%
\pagestyle{plain}
The property of space-time has fundamental importance in
understanding the law of nature. Noncommutative (NC) quantum
field theory, which modifies the space-time conmmutation relations, 
provides an alternative to the ordinary quantum field
theory which may shed some light on the detailed structure of
space-time. A simple way to modify the space-time properties 
is to change the usual space-time coordinate $x$ to 
nonconmmutative coordinate $\hat X$ such that\cite{1}

\begin{eqnarray}
[\hat X^\mu, \hat X^\nu] = i \theta^{\mu\nu}.
\label{eq:NCgeo}
\end{eqnarray}
We consider the case with $\theta^{\mu\nu}$ be a constant 
real anti-symmetric matrix which commutes with $\hat X^\mu$. 

NC quantum field theory based on the above commutation relation 
can be easily studied using the Weyl-Moyal 
correspondence replacing the product of two fields $A(\hat X)$ and $B(\hat X)$
with NC coordinates by the star ``*'' product\cite{2},

\begin{eqnarray}
A(\hat X) B(\hat X) \to A(x)*B(x) = Exp[i
{1\over 2} \theta^{\mu\nu}\partial_{x,\mu}
\partial_{y,\nu}] A(x) B(y)|_{x=y}.
\end{eqnarray}

Phenomenology for NC electromagnetic theory $U(1)_{em}$ has been
studied recently\cite{3}, but less has been done for noncommutative
Standard Model (NCSM) because it is non-trivial to construct such a theory.
Due to the noncommuting nature of the ``*'' product, even with a $U(1)$
gauge theory the charges of matter fields in the theory 
are fixed to only three possible values, 1, 0, -1\cite{4}. Also $SU(N)$
group can not simply be gauged with ``*'' product, but $U(N)$ can be\cite{4}. 
These pose certain 
difficulties in constructing NCSM, since
in the SM the $U(1)_Y$ charges are not just 1, 0, -1, but some of them are
fractionally charged, such as 1/6, 2/3, -1/3 for left handed quarks, right
handed up and down quarks, respectively. There are also the problems
with gauging the $SU(3)_C$ and $SU(2)_L$ groups. 
However, all these difficulties
can be overcome with the use of the Seiberg-Witten map\cite{5} which maps
noncommutative $U(N)$ gauge fields to commutative $SU(N)$ gauge fields. 
Therefore, consistent noncommutative $SU(N)$ gauge theory can be constructed.
The same map can also cure the charge quantization problem as shown in Ref.
\cite{6,7}
by introducing new degrees of freedom.
With the help of Seiberg-Witten map 
specific method to construct NCSM has been developed\cite{6,7}.

To solve the $U(1)$ 
charge quantization problem, one associates each charge $gq^{(n)}$ of
the nth matter field a gauge field $\hat 
a_\mu^{(n)}$\cite{7}. In the commutative limit,
$\theta^{\mu\nu} \to 0$, $\hat 
a_\mu^{(n)}$ becomes the single gauge field $a_\mu$ of the ordinary 
conmmuting space-time $U(1)$ gauge theory. 
But at non-zero orders in $\theta^{\mu\nu}$, $\hat a^{(n)}_\mu$ receives corrections
and becomes\cite{7}

\begin{eqnarray}
\hat a^{(n)}_\xi 
= a_\xi + {gq^{(n)}\over 4} \theta^{\mu\nu}\{a_\nu, \partial_\mu
a_\xi \} + {gq^{(n)}\over 4} \theta^{\mu\nu} \{f_{\mu\xi}, a_\nu\} +
O(\theta^2),
\end{eqnarray}
where $f_{\mu\nu}$ is the field strength of $a_\mu$. In doing so, 
the kinetic energy of the gauge boson will, however,  
be affected. Depending on
how the kinetic energy is defined (weight over different field strength of
$\hat a_\mu^{(n)}$), the resulting 
kinetic energy will be different even though 
the proper normalization to obtain
the correct kinetic energy in the commutative limit is imposed\cite{7}.

In the Standard Model, there are six 
different matter field multiplets for each
generation, i.e. $U_R$, $d_R$, $(u,\;d)_L$, $e_R$, $(\nu,\;e)_L$ and
$(H^0,\;H^-)$,
a priori one can choose 
a different 
$g_i$ for each of them. After identifying three combinations with the
usual $g_3$, $g_2$ and $g_1$ couplings for the SM gauge groups, there is
still freedom to choose different gauge boson self interaction couplings 
at non-zero orders in $\theta^{\mu\nu}$ 
(we refer these choices as different schemes). This leads to
ambiguities in self interactions of gauge bosons when non-zero order terms in 
$\theta^{\mu\nu}$ are included. This point was nicely
demonstrated in Ref.\cite{7}.   
It is shown there that with one particular scheme,
there are no triple photon self interactions, 
and with a different scheme
there are. This results in non-uniqueness of the theory. 
This problem needs to
be resolved either by performing experiments to test different schemes or 
applying a underlying theory which uniquely determines the 
kinetic energy terms. In this paper we
propose a solution to this problem from grand unification theory point
of view.

The problem with the non-uniqueness of gauge boson kinetic energy is due to
the fact that in order to overcome the charge quantization problem, 
new degrees of freedom have to be introduced. If the degrees of freedom can be 
reduced and at the same time 
the correct charge quantization can be obtained, the problem
will be solved. To this end we note that grand unified
theory with a single gauge coupling does exactly this. Kinetic energy for
SM fields deduced from such a unified theory will be uniquely determined.
We will work with $SU(5)$ grand unification\cite{8} to demonstrate this and 
concentrate on triple gauge boson interactions among the SM fields, photon
$\gamma$, $Z$ boson and gluons $G$. 

Using Seiberg-Witten map, one can easily obtain the kinetic energy of 
a noncommutative $SU(5)$ gauge theory. 
The noncommutative $SU(N)$ gauge boson 
interactions, when
written in terms of commutative conventional gauge fields $A^0_\mu$ and its 
field strength $F^0_{\mu\nu} = \partial_\mu A^0_\nu - \partial_\nu A^0_\mu
-i g_N [A^0_\mu, A^0_\nu]$
to the leading order in
$\theta^{\mu\nu}$, are given by\cite{6,9}

\begin{eqnarray}
L = -{1\over 2} Tr (F^0_{\mu\nu}F^{0\mu\nu})
+ g_N\theta^{\mu\nu}[{1\over 4}Tr(F^0_{\mu\nu}F^0_{\rho\sigma}F^{0\rho\sigma})
-Tr(F^0_{\mu\rho}F^0_{\nu\sigma}F^{0\rho\sigma})].
\label{ss}
\end{eqnarray}

The above Lagrangian is uniquely determined to order $\theta$. 
If a low energy theory is deduced from such a theory, then all
the gauge couplings are completely fixed. We now use
the noncommutative $SU(5)$ self gauge interactions to derive
the NCSM triple neutral gauge boson couplings.

With appropriate Higgs boson representations and corresponding 
non-zero vacuum expectation values (VEV), such as a 24 and a 5,
$SU(5)$ gauge group can be broken down to the SM 
gauge group, $SU(3)_C\times
SU(2)_L\times U(1)_Y$, and then to $SU(3)_C\times U(1)_{em}$. 
The SM gauge boson self interactions
can be obtained by integrating out the heavy degrees of freedom.

The gauge bosons in $SU(5)$ theory are contained in the 24 adjoint 
representation. In addition to the 8 color gauge bosons $G^a$, 
4 electroweak gauge bosons, $W^\pm$, $Z$ and $\gamma$, there are also 12
colored heavy bosons, the $X$ and $Y$ bosons. This representation,
in terms of the gauge fields, can be written as\cite{8}

\begin{eqnarray}
&&\sqrt{2}A^0_\mu = \sqrt{2}T^a A^{0a}_\mu = \nonumber\\
&&\left (
\footnotesize{ \begin{array}{lllll}
G^{11}_\mu -2B_\mu/\sqrt{30}& G^{12}_\mu&G^{13}_\mu& \bar X^1_\mu&
\bar Y^1_\mu\\
G^{21}_\mu&G^{22}_\mu -2B_\mu/\sqrt{30}&G^{23}_\mu&\bar X^2_\mu&
\bar Y^2_\mu\\
G^{31}_\mu&G_{32}&G^{33}_\mu -2B_\mu/\sqrt{30}&\bar X^3_\mu&
\bar Y^3_\mu\\
X^1_\mu&X^2_\mu&X^3_\mu&W^3_\mu/\sqrt{2}+3B_\mu/\sqrt{30}&W_\mu^+\\
Y^1_\mu&Y^2_\mu&Y_\mu^3&W_\mu^-&-W^3_\mu/\sqrt{2}+3B_\mu/\sqrt{30}
\end{array} }
\right ),
\end{eqnarray}
where $T^a$ are the $SU(5)$ generators with normalization
$Tr(T^aT^b) = \delta^{ab}/2$. $G^{ij}_\mu$ correspond to SM gluon fields,
and $G^{ii}$ are linear combinations of the usual $G^3$ and $G^8$ gluons.
Linear combinations of $W^3_\mu$ and $B_\mu$ give rise to the
$\gamma$ and $Z$ fields.
We will indicate the generators of $B_\mu$, 
$W^3_\mu$, and $G^a_\mu$ fields by $T^B$, $T^{w3}$ and $T^{Ga}$, respectively.

To obtain the triple neutral gauge boson self interactions, one needs to 
expand the $Tr(F^0F^0F^0)$ terms in eq. (\ref{ss}).  There are only four
types of triple 
neutral gauge boson interactions, $BBB$, $W^3W^3B$, $GGB$ and $GGG$
resulting from $Tr(F^0F^0F^0)$ at the order $\theta_{\mu\nu}$. 
The corresponding
traces needed to evaluate the triple neutral gauge boson interactions
are 

\begin{eqnarray}
&&Tr(T^BT^BT^B) = {1\over 4 \sqrt{15}},\;\;\;\;
Tr(T^{w3}T^{w3}T^B) = {3\over 4\sqrt{15}},\nonumber\\
&&Tr(T^{Ga}T^{Gb}T^B) = - {1\over 2 \sqrt{15}}\delta^{Ga\;Gb},\;\;\;\;
Tr(T^{Ga}T^{Gb}T^{Gc}) = {1\over 4} (i f^{GaGbGc} +d^{GaGbGc}),
\end{eqnarray}
where $f^{abc}$ and $d^{abc}$ are the totally anti-symmetric and symmetric
structure functions for the color gauge $SU(3)_C$ group.

Using the above, we obtain the triple neutral gauge boson interactions as

\begin{eqnarray}
L &=& 
\theta^{\mu\nu}g_5{1\over 4 \sqrt{15}} [ {1\over 4}
B_{\mu\nu}B_{\rho\sigma} - B_{\mu\rho}B_{\nu\sigma}]B^{\rho\sigma} \nonumber\\
&+& \theta^{\mu\nu} g_5 {3\over 4 \sqrt{15}}
[({1\over 4}W^3_{\mu\nu} W^{3}_{\rho\sigma} - W^3_{\mu\rho}W^3_{\nu\sigma})
B^{\rho\sigma}\nonumber\\ 
&+& ({1\over 4} W^3_{\mu\nu} B_{\rho\sigma} - W^3_{\mu\rho}B_{\nu\sigma})
W^{3,\rho\sigma} +
({1 \over 4}B_{\mu\nu} W^3_{\rho\sigma} - B_{\mu\rho}W^3_{\nu\sigma})
W^{3,\rho\sigma}]\nonumber\\
&-&\theta^{\mu\nu}g_5{1\over 2\sqrt{15}}
[({1\over 4}G^a_{\mu\nu} G^{a}_{\rho\sigma} - G^a_{\mu\rho}G^a_{\nu\sigma})
B^{\rho\sigma}\nonumber\\
&+&({1 \over 4} G^a_{\mu\nu} B_{\rho\sigma} - G^a_{\mu\rho} B_{\nu\sigma})
G^{a,\rho\sigma} + 
({1\over 4}B_{\mu\nu} G^a_{\rho\sigma} - B_{\mu\rho}G^a_{\nu\sigma})
G^{a,\rho\sigma}]\nonumber\\
&+& \theta^{\mu\nu} g_5 {1\over 4} d^{abc}
({1\over 4} G^a_{\mu\nu}G^b_{\rho\sigma}G^{c,\rho\sigma}
-G^{a}_{\mu\rho}G^b_{\nu\sigma}G^{c,\rho\sigma}).
\end{eqnarray}

When the gauge 
group is further broken down to $SU(3)_C\times U(1)_{em}$ via non-zero
VEV of the 5, one obtains 
the new triple  gauge couplings in the physical fields $\gamma (A)$, $Z$ and 
$G$.
We have,

\begin{eqnarray}
L &=& 
{g_5\over 4\sqrt{15}} \theta^{\mu\nu}
 \left \{
c(\theta) (5-4c(2\theta))
\left [ {1\over 4}F_{\mu\nu} F_{\rho\sigma}F^{\rho\sigma} 
- F_{\mu\rho}F_{\nu\sigma} 
F^{\rho\sigma} 
\right]\right.\nonumber\\
&+&s(\theta)(1+4 c(2\theta))
\left [{1\over 4} (2 F_{\mu\nu} Z_{\rho\sigma}F^{\rho\sigma}
+Z_{\mu\nu}F_{\rho\sigma}F^{\rho\sigma}) 
-(2 F_{\mu\rho}Z_{\nu\sigma}
F^{\rho\sigma} + F_{\mu\rho}F_{\nu\sigma}Z^{\rho\sigma})
\right ]\nonumber\\
&-& 
c(\theta)(1-4c(2\theta))
\left [ {1\over 4}(2 Z_{\mu\nu} Z_{\rho\sigma}F^{\rho\sigma}
+F_{\mu\nu}Z_{\rho\sigma}Z^{\rho\sigma})
-( 2 Z_{\mu\rho} 
F_{\nu\sigma}Z^{\rho\sigma} + Z_{\mu\rho}Z_{\nu\sigma}F^{\rho\sigma})
\right ]\nonumber\\
&-&s(\theta)(5+4c(2\theta)
\left .\left [ {1\over 4} Z_{\mu\nu}Z_{\rho\sigma}Z^{\rho\sigma}
- Z_{\mu\rho}Z_{\nu\sigma}Z^{\rho\sigma}
\right ]\right \} \nonumber\\
&-& {g_5\over 8\sqrt{15}} \theta^{\mu\nu}
\left [ 2G_{\mu\nu}^aG^a_{\rho\sigma}(c(\theta) F^{\rho\sigma}
-s(\theta) Z^{\rho\sigma})
+ (c(\theta) F_{\mu\nu} - s(\theta) Z_{\mu\nu}) G^a_{\rho\sigma} 
G^{a,\rho\sigma}\right .\nonumber\\
&-&\left . 4G_{\mu\rho}^a G_{\nu\sigma}^a ( c(\theta) F^{\rho\sigma} - s(\theta)
Z^{\rho\sigma}) - 8(c(\theta) F_{\mu\nu} - s(\theta) Z_{\mu\nu}) 
G^a_{\rho\sigma}G^{a,\rho\sigma}\right ]\nonumber\\
&+& {g_5\over 4} \theta^{\mu\nu} d^{abc}
({1\over 4} G^a_{\mu\nu}G^b_{\rho\sigma}G^{c,\rho\sigma}
-G^{a}_{\mu\rho}G^b_{\nu\sigma}G^{c,\rho\sigma}).
\end{eqnarray}
where $c(n\theta) = \cos(n\theta_W)$, $s(n\theta) = \sin(n\theta_W)$, and 
$F_{\mu\nu} = \partial_\mu A_\nu - \partial_\nu A_\mu$, 
$Z_{\mu\nu} = \partial_\mu Z_\nu - \partial_\nu Z_\mu$.
Note that each of the coupling and $\sin\theta_W$ given above are at the unification scale. They have to be evolved to the appropriate 
energy scale depending on the experiment.

It is clear that the triple 
neutral gauge boson self interactions in the NCSM obtained above 
are completely specified. 
The $\gamma\gamma\gamma$, $\gamma\gamma Z$, $\gamma ZZ$, $ZZZ$,
$\gamma GG$, $ZGG$ and $GGG$ types of interactions are all new compared with
the SM predictions.
The corresponding couplings are not zero 
which will lead to observational effects if the energy scale of the
nonconmmutativity is not too high (less than a few TeV)\cite{3}. 

The triple photon interaction can be studied in various processes, such as
$e^+e^-\to \gamma\gamma$, $e\gamma \to e\gamma$ and $\gamma\gamma \to e^+e^-$.
The analyses can be carried out in similar ways as those discussed in
Ref.\cite{3} with appropriate modifications of the vertices involved.
The sensitivity to the nonconmmutative parameter $\theta^{\mu\nu}$
is typically in the TeV range at next generation of linear colliders 
with center of mass energy around one TeV\cite{3}.
The $\gamma \gamma Z$ coupling can also be studied in the above processes,
in $e^+e^- \to ZZ,\; \gamma Z$, $Z\to \gamma \gamma$, 
and on shell production at $\gamma\gamma \to Z$
processes. 
The $\gamma ZZ$ coupling can be studied in the processes $e^+e^- \to ZZ,\;
\gamma Z$. The couplings involving gluons can also be studied at $e^+e^-$
colliders, such as $e^+e^- \to GG$, and in hadron colliders, such as
$pp(\bar p) \to GG$, 
and in $Z\to GG$ decay. All the processes having gluons
in the final state may be more difficult to study compared with the ones
without due to hadronization. 
But if sophisticated polarization analyses can be carried out,
these processes can also provide interesting information about the
nonconmmutativity of space-time. 

In conclusion, in this paper we have proposed a solution to the
arbitrariness of kinetic energy problem of the noncommutative 
Standard Model from grand unification point of view. The kinetic energy 
and the resulting self gauge boson interactions are uniquely determined
in noncommutative $SU(5)$ theories. 
Although we have only worked with $SU(5)$ grand unification
as an example, the conclusion will hold in any grand unification theories 
with a single coupling constant. At the first order in the noncommutative
parameter $\theta^{\mu\nu}$, there are new triple gauge boson interactions,
such as 
$\gamma\gamma\gamma$, $\gamma\gamma Z$, $\gamma ZZ$, $ZZZ$, $\gamma GG$,
$ZGG$ and $GGG$.  These interaction
are forbidden in the Standard Model. The theory can be tested by experiments.
Non observation of these interactions can also put bounds on the 
noncommutative parameter $\theta^{\mu\nu}$.

We thank P.-M. Ho for discussions.
This work was supported in part by DOE grant DE-FG03-96ER40969 and 
by National Science Council under grants NSC 89-2112-M-002-058 and NSC
89-2112-M-002-065, and in part
by the Ministry of Education Academic Excellence Project 89-N-FA01-1-4-3.
One of us (NGD) would like to thank hospitality provided by the 
Department of Physics at National Taiwan University where this work was done.

\end{document}